\documentstyle[twocolumn,epsf,aps]{revtex}
\draft
\begin{document}
\title{Anisotropic Ordering in a Two-Temperature Lattice Gas}
\author{Attila Szolnoki and Gy\"orgy Szab\'o}
\address
{Research Institute for Materials Science, H-1525 Budapest, POB 49, Hungary}
\address{ \centering{ \ medskip\em
\begin{minipage}{14cm}
{}~~~We consider a two-dimensional lattice gas model with repulsive nearest-
and
next-nearest-neighbor interactions that evolves in time according to
anisotropic
Kawasaki dynamics. The hopping of particles along the principal directions is
governed
by heat baths at different temperatures.
 The stationary states of this non-equilibrium
model are studied using a generalized dynamical mean-field analysis. In the
stable state the particles form parallel chains which are oriented along the
direction of the higher temperature. The resulting phase diagram is
confirmed by Monte Carlo simulations.
\end{minipage}
}}

\maketitle

\narrowtext

Systems driven into a non-equilibrium steady state can reveal a number of
features not characteristic to the equilibrium state. An example is the
well-known
two-dimensional lattice gas model with attractive
nearest-neighbor interaction whose degenerate ground states violate the
$x-y$ symmetry of the system. Below a critical temperature the disordered (high
temperature) phase segregates
into a high- and a low-density phase and the strip region
of the condensed particles can be oriented either horizontally or vertically if
periodic boundary conditions are imposed. There are several
different ways to provide non-equilibrium conditions and to
violate the
$x-y$ symmetry. In driven lattice gases introduced by Katz {\it et al.}
\cite{kls}
the particle jumps are under the influence of a uniform external electric field
whose direction is chosen to be parallel to one of the principal axes
\cite{zia}.
Another way to define anisotropic dynamics is the so called two-temperature
models
are investigated extensively
\cite{cheng,danzia,racz,bass}. In these former models the particle jumps are
coupled
to two different heat baths, that is, the Kawasaki (exchange) dynamics is
characterized by $T_x$ and $T_y$ temperatures along the $x$ and $y$ axes. In
contrary to the previous dynamics, here there is no a uniform particle current
through the system. At the same time, an energy transport exists from one of
the
heat baths to the other one driving the system into a non-equilibrium steady
state.
This approach can describe the effect of an alternating field (or polarized
light)
on the ordering process \cite{zia}.

In the attractive interacting system the symmetry breaking is represented by
the
orientation of
the interface separating the high- and a low-density phases. Significantly
different symmetry breaking can be observed in the model discussed by Sadiq
and Binder \cite{sadiq} which is characterized by repulsive nearest- and
next-nearest-neighbor interactions with equal strength. In the half-filled
system
the particles form alternately occupied and empty columns (or rows). That is,
here
the symmetry breaking appears directly (as a bulk feature) in the fourfold
degenerate ground states. The ordering process in this model is well understood
\cite{sadiq,SB,ole} and the effect of the uniform driving field on the ordering
process
was also investigated \cite{sz1}. It was found that the phases consisting of
chains
perpendicular to the external field become unstable and the stationary state is
oriented along the field direction. In other words, the chain
orientation of the ordered phase can be controlled by the application of an
electric
field with suitable direction. Similar effect is suspected when using an
alternating
field. This conjecture was the motivation of the present work. Now
we will study the effect of anisotropic jump rate on the ordering process
within the
formalism of the two-temperature model.

First we introduce the model and investigate by adopting a simple dynamical
mean-field
theory
\cite{sz1}. Linear stability analysis has been performed to distinguish the
stable
and unstable solutions. In the light of these results the phase diagram is
evaluated at higher level (4-point approximation) of dynamical mean-field
theory. Finally the
results are compared with Monte Carlo (MC) data.

We consider a half-filled, lattice-gas model with isotropic nearest- and
next-nearest-neighbor repulsive interactions with equal strength on a square
lattice. As usual the coupling constants as well as the Boltzmann constant are
chosen
to be unity. The kinetics is governed by Kawasaki dynamics characterized by
single
particle jumps to one of the empty nearest neighbor sites \cite{kaw}. To avoid
the difficulties arising from the non-analytic feature of Metropolis rate,
Kawasaki hopping rate is used for directions $\alpha = x$ or $y$:
\begin{equation}
g_{\alpha}(\Delta  H) =  {1 \over { 1 + \exp(\Delta H / T_{\alpha}) }} \ \ ,
\label{eq:hoppprob}
\end{equation}
where $\Delta H$ is the energy difference between the final and initial
configurations. This jump rate is anisotropic and satisfies detailed balance at
temperature $T_y$
($T_x$) in the vertical (horizontal) direction.

The equilibrium model ($T_x = T_y$) undergoes a continuous phase transition at
a
critical temperature $T_c= 0.525$. In the ground states the columns (or rows)
are alternately occupied and empty providing fourfold degeneracy. Following the
notation
of Sadiq and Binder the lattice is divided into four interpenetrating
sublattices
and the $2 \times 1$ or $1 \times 2$ long-range order is characterized by the
corresponding average sublattice occupations $n_i$ ($i=1,\ldots,4$)
\cite{sadiq}.
In the half-filled system the $2 \times 1$ states consisting of vertical chains
are given as $n_1 = n_4 =(1+m_x)/2$ and $n_2=n_3=(1-m_x)/2$, where
$-1 \le m_x \le 1$ is the order parameter. Notice that $m_x$ can be either
positive
or negative. Its absolute value as a function of temperature may be estimated
by
using the traditional mean-field approximation. Similar expressions are
obtained
for the $1 \times 2$ states of horizontal chains, namely $n_1=n_2=(1+m_y)/2$
and $n_3=n_4=(1-m_y)/2$. It is emphasized that the four possible states are
equivalent ($|m_x|=|m_y|$) in the
equilibrium system. Evidently, in the high temperature disordered phase
$m_{\alpha}=0$.

In the following simple dynamical mean-field approximation \cite{sz1} we will
suppose that the long-range order still exits under the present non-equilibrium
condition. It means that the anisotropic jump rate results in only a
difference between the solutions characterized by $m_x$ and $m_y$.

According to Eq.~(\ref{eq:hoppprob}) the time evolution of $n_1$ is given by
summarizing the contributions of the jumps modifying the particle density
in sublattice 1 as
\begin{eqnarray}
{dn_1 \over dt} &=& - 2 n_1[(1- n_2) g_x(E_{21}) + (1-n_4) g_y(E_{41})]
\nonumber \\
& & + 2 (1-n_1)[n_2 g_x(E_{12}) + n_4 g_y(E_{14})] \,\,\, ,
\label{eq:master}
\end{eqnarray}
where $E_{ij}$ is the average value of $\Delta H$ when a particle jumps from
sublattice
$i$ to $j$. Similar equations can be derived for the rest of sublattice
occupations.
Assuming a $2 \times 1$ or $1 \times 2$ structures
(vertical or horizontal chains) the stationary solution of
Eq.~(\ref{eq:master})
obeys the following general form:
\begin{equation}
{{1+m_{\alpha}} \over {1-m_{\alpha}}} = \exp(3m_{\alpha} / 2T_{\alpha})
\label{eq:implicit}
\end{equation}
where $\alpha=x$ or $y$ respectively. Notice that the jumps along the chains do
not modify the particle densities within the chains therefore the value of the
corresponding order parameter depends only on the perpendicular jump rate (or
perpendicular temperature). The order parameter as a function of perpendicular
temperature can be determined numerically. This simple approximation predicts
continuous phase transition with a critical temperature $T_c^{(MF)}=3/4$.

\begin{figure}
\centerline{\epsfxsize=8cm
                   \epsfbox{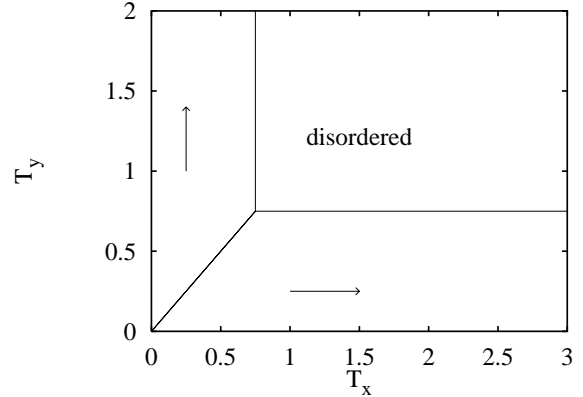}
                   \vspace*{2mm}     }
\caption{Phase diagram predicted by the simplest mean-field theory. The arrows
indicate
the stable chain directions in the ordered regions.}
\label{fig:MFpd}
\end{figure}

Obviously, we have two different solutions if $T_x \ne T_y$. In this situation
the linear stability analysis is used to determine which solution remains
stable.
This method proved to be very useful when investigating the effect of the
uniform
driving field on these ordered structure in the same lattice-gas model
\cite{sz1}.
Now we restrict ourselves to the survey of results without dealing with
mathematical
details given in a previous work \cite{sz1}. It is found that the ordered
structure
appears when $ \min (T_x, T_y) < T_c^{(MF)}$ and the chains in the stable
phases are
parallel to direction belonging to the higher temperature. As mentioned above,
the
value of the corresponding order parameter depends only on the perpendicular
temperature.
As a consequence the chain structure remains stable even in the limit case when
the
parallel temperature goes to infinity. The resultant phase diagram is
illustrated in Fig.~\ref{fig:MFpd}.

The above results imply the possibility of reorientation in the low temperature
region.
Namely, exchanging the values of $T_x$ and $T_y$ a particle rearrangement is
expected
as it is observed by MC simulations under the effect of a uniform field
\cite{sz1}. It
is rather difficult to realize the two-temperature model, but as we mensioned
earlier
an alternate electric field can cause similar behavior.
Detailed analysis of the reorientation phenomenon goes beyond the scope of the
present
work.
Instead, we focus our attention to have a more accurate phase diagram.

The prediction of the above simple mean-field theory can be improved by using
the
generalized mean-field theory at the levels of 2- and 4-point approximations.
In these calculations we have to determine the probability of all the possible
configurations on 2- and 4-point clusters by evaluating numerically the
stationary
solution of a set of equations which describe the time variation of
configuration
probabilities. Details of such a calculation are given in previous papers
\cite{dickm,szolszab}.
The 2-point (pair) approximation fails to reproduce the phase transition even
in the
equilibrium model ($T_y = T_x$). More precisely, this approximation predicts
first-order transition. Similar discrepancies have already been observed for
other sytems \cite{dickm,CA}. This shortage of the present technique may be
corrected by calculating with larger cluster(s).

Exploiting symmetries we have (only) four independent parameters to describe
the
probability of all the possible particle configurations on a square at the
level of
4-point approximation. In equilibrium this method suggests a
continuous order-disorder transition and the predicted critical temperature is
$T_c^{(4p)}=0.566$ which is close to the exact value mentioned above.

\begin{figure}
\centerline{\epsfxsize=8cm
                   \epsfbox{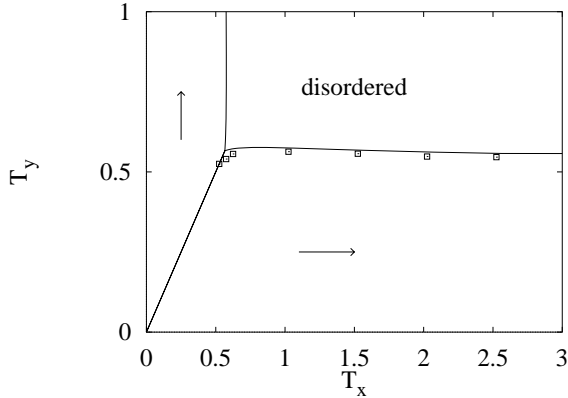}
                   \vspace*{2mm}     }
\caption{Comparison of phase diagrams predicted by 4-point approximation (solid
line) and MC simulations (squares).}
\label{fig:pd}
\end{figure}

In the non-equilibrium case ($T_x \ne T_y$) the prediction of the 4-point
approximation
agrees qualitatively with those obtained above (see Fig.~\ref{fig:MFpd}).
However,
now the critical temperature depends on both temperature. The numerical results
are summarized in a revised phase diagram as plotted in Fig.~\ref{fig:pd}. A
weak anomaly
has been found when $T_x$ and $T_y$ are close to $T_c^{(4p)}$. Namely,
the ordering temperatures {\it increase} first when moving away the
equilibrium. After
a low peak
the transition temperature decreases toward a constant value,
$T_c^{(4p)}(\infty)=
0.94 T_c^{(4p)}$ if the parallel temperature goes to infinity. Hence, in
agreement
with the earlier approximation, the ordered state still exists even at infinite
parallel temperature if the perpendicular temperature is low enough.

To check the above  predictions a series of Monte Carlo simulations \cite{oleb}
has been carried out
varying the temperatures $T_x$ and $T_y$. Using periodic boundary conditions
the simulations
are performed on a square system for different sizes
$L = 20$, 30, 40, 80, and 200. The size dependence of data has indicated no
serious finite size effects.
We have determined the time averages of the energy and order parameter varying
the perpendicular temperature. These simulations have justified that the
ordering
process remains continuous for all the parallel temperatures we studied. From
these data we
could determine the value of critical temperatures. The results confirm the
above
theoretical predictions as demonstrated in Fig.~\ref{fig:pd}. This figure
vizualizes that the transition temperature depends sligthly on the parallel
temperature, in a very similar way as suggested by the 4-point approximation
and its
value tends to $0.98 T_c^{(MC)}$ when $T_x$ or $T_y$ goes to infinity.

In summary, we have studied the non-equilibrium ordering process in a
two-dimensional
lattice gas coupled to two thermal baths at different temperatures. In the
equilibrium
system the particles form parallel chains oriented either horizontally or
vertically.
Using a simple mean-field theory including linear stability analysis we have
found that
the ordered structure appears in the nonequilibrium case if the lower
temperature is
less than a critical value and the chains are oriented along the higher
temperature.
This picture has been confirmed by dynamical 4-point approximation and MC
simulations.
Despite the prediction of the simple mean-field theory the sophisticated
methods
have indicated that the transition temperature (as well as the temperature
dependence of order parameter) depends slightly on the parallel temperature.

The possibility of orientation (or reorientation) process may appear in other
two-dimensional systems belonging to the family of Ashkin-Teller models
\cite{AT}.
The present two-temperature approach gives some insight into this phenomena
even for higher dimensions.

\vspace{1cm}

The authors are grateful to Christof Scheele for careful reading of the
manuscript.
This research was supported by the Hungarian National Research Fund (OTKA)
under
Grant No. F-7240.

\end{document}